\documentclass{egpubl}
\usepackage{envirvis2022}

\WsPaper         %
\electronicVersion %

\ifpdf \usepackage[pdftex]{graphicx} \pdfcompresslevel=9
\else \usepackage[dvips]{graphicx} \fi
\usepackage{lineno}

\usepackage{t1enc,dfadobe}

\usepackage{hyperref, egweblnk}
\usepackage{cite}

\title[Towards a Survey of Visualization Methods for Power Grids]%
{Towards a Survey of Visualization Methods for Power Grids}

\author[submission ID]{TODO}

\author[Fischer et al.]
       {Maximilian. T. Fischer$^{1}$ \orcid{0000-0001-8076-1376}
        and Daniel A. Keim$^{1}$ \orcid{0000-0001-7966-9740}\\
	         $^1$University of Konstanz, Germany\\
	         }

\usepackage{wasysym}
\usepackage{amssymb}
\newcommand{\symYes}{\CIRCLE}
\newcommand{\symNo}{\Circle}
\newcommand{\symPartial}{\LEFTcircle}
\newcommand{\symLow}{{\fontsize{4}{6}\selectfont $\blacksquare~\square~\square$}}
\newcommand{\symMedium}{{\fontsize{4}{6}\selectfont $\blacksquare~\blacksquare~\square$}}
\newcommand{\symHigh}{{\fontsize{4}{6}\selectfont $\blacksquare~\blacksquare~\blacksquare$}}

\usepackage[table,xcdraw,pdftex,dvipsnames]{xcolor}
\usepackage{colortbl}
\usepackage{array}
\usepackage{booktabs}
\usepackage{graphics}
\usepackage{graphicx}
\definecolor{ColorBGTableComplextiy}{HTML}{EDF4FB}
\definecolor{ColorBGTableVisualization}{HTML}{FCE5E5}
\definecolor{ColorBGTableProperties}{HTML}{F1F7E8}
\definecolor{ColorBGTableStudy}{HTML}{FDF3E5}
\newcolumntype{A}{>{\columncolor{ColorBGTableComplextiy!50}}c}
\newcolumntype{B}{>{\columncolor{ColorBGTableVisualization!50}}c}
\newcolumntype{C}{>{\columncolor{ColorBGTableProperties!50}}c}
\newcolumntype{D}{>{\columncolor{ColorBGTableStudy!50}}c}

\definecolor{tableblue}{HTML}{709BBF}
\definecolor{theader}{HTML}{6DAFFF}
\definecolor{tablegrid}{HTML}{B6CBDC}
\definecolor{twhite}{HTML}{FFFFFF}
\definecolor{tlightgrey}{HTML}{EEEEEE}
\definecolor{tblack}{HTML}{000000}
\definecolor{tdarkorange}{HTML}{F56B00}
\definecolor{torangelight}{HTML}{FFCE93}
\definecolor{tmandarin}{HTML}{FFC702}
\definecolor{tmandarinlight}{HTML}{FFD7C2}

\definecolor{tred}{RGB}{255,59,48}
\definecolor{tredl}{RGB}{255,177,172}
\definecolor{torange}{RGB}{255,149,0}
\definecolor{torangel}{RGB}{255,213,153}
\definecolor{tyellow}{RGB}{255,204,0}
\definecolor{tyellowl}{RGB}{255,235,153}
\definecolor{tgreen}{RGB}{76,217,100}
\definecolor{tgreenl}{RGB}{183,240,193}
\definecolor{ttealblue}{RGB}{90,200,250}
\definecolor{ttealbluel}{RGB}{189,233,253}
\definecolor{tblue}{RGB}{0,122,255}
\definecolor{tbluel}{RGB}{153,202,255}
\definecolor{tpurple}{RGB}{8,86,214}
\definecolor{tpurplel}{RGB}{156,187,239}
\definecolor{tpink}{RGB}{255,45,85}
\definecolor{tpinkl}{RGB}{255,171,187}

\begin{document}

	\maketitle
	
	\begin{abstract}
		With the ongoing emergence of smart and distributed grids, it becomes increasingly important to understand as well as improve legacy infrastructure while operating a much more interconnected and fragile architecture.
		To support this endeavor, a detailed simulation and real-life analysis are required.
		Leveraging advanced visualization and analytics methods can significantly improve and simplify tasks such as network analysis, maintenance, and planning, while also enabling operators to spot critical issues which are hard to detect otherwise.
		In this work, we work towards a comprehensive overview of the methods developed for the interactive visualization of power grids.
		We give an overview of the development of the field before motivating a range of comparison criteria and then evaluating the advantages and disadvantages of the single approaches.
		Finally, we derive a set of open research questions and possible further improvements to the field.
	\end{abstract}

	\section{Introduction}
	The awareness of the geographic context is an important factor for many applications incorporating network data, like power, water supply, road, and fiber networks.
	Visualizing these networks on a geographical map is challenging since nodes have a "fixed" position, and thus clutter or occlusion may be harder to avoid, as techniques like edge bundling, clustering, aggregations, or animations have to be applied with care~\cite{Becker.1995}.
	In this work, we analyze how the progress in different visualization and visual analytics techniques in hugely complex multivariate, multi-temporal, and spatial data sets ~\cite{Keim.2001,Keim.2002,Jin.2009, Oliveira.2003} has taken hold in the visualization of power grids~\cite{Nga.2012}.
	
	A \emph{power grid} can be described as an interconnected network of electrical transmission lines and associated equipment distributing electricity from electricity sources to consumers.
	With the ongoing emergence of smart and distributed grids, it becomes increasingly important to understand and improve legacy infrastructure while operating a much more interconnected and fragile architecture~\cite{Overbye.2014}.
	Visualization and simulation are essential to understand the capabilities and limits of a given network, identify and resolve specific issues, and optimize the network structure to adapt to new and more complex usage scenarios.
	When visualizing a power grid, two different aspects are often shown separately:
	power data and the geographical transmission grid.
	Basic visualization methods like line diagrams, histograms, and bar charts are commonly used to visualize power grid data and are helpful for basic tasks.
	The transmission grid is often visualized in a topological, approximated geographical or stylized form.
	Both forms are rarely combined, but could result in new insights, as the grid and its associated data share a complex interdependence.
	Using information mining to discover such relationships in the data and visualize the results in relation to the grid structure allows these interdependencies to be presented in new and useful ways to gain a more profound and faster understanding of the data presented.
	Visualization methods in power grid control centers employed in the last century often were simple.
	Almost all operational quantities like voltages, status indicators, and power flow used to be represented either analog on the vertex elements or as numeric values on table displays.
	System information on dynamic displays is often very limited, for example, the use of font color to display quantity limits or the use of dashed lines for device status~\cite{Overbye.2007}.
	The whole system is often shown on a static wall map for an overview, and different colored lights are the only dynamic data displayed~\cite{Overbye.2007}.
	Nowadays, advanced visualization techniques, with the emergence of wide-area displays, have become slightly more common in power control centers.
	Examples include colored contour plots to display the difference in voltages across large network regions, moving arrows to show line flow, and dynamically resized charts for grid elements reaching their operational limit or are out of service. However, much room for improvement still exists~\cite{Overbye.2007}.
	Applications that benefit from such visualizations include network monitoring, dynamic load management, fault detection, network planning, and scenario analysis.
	
	While a few overview articles and surveys exists~\cite{Nga.2012, MikkelsenC.2012, rao.cn.2004, Cuffe.2015, Poursharif.2015a}, they are several years old and none is collectively exhaustive.
	This survey tries to bridge the gap and provide a comprehensive overview of the latest, non-trivial methods developed to (interactively) visualize power grids and combine both power data and grid representation in one visualization.
	Thereby, we make the following contributions:
	\begin{itemize}
	    \item An overview of the development of the field and related work
	    \item A comparative evaluation and discussion of the approaches
	\end{itemize}
	Finally, we discuss key findings and pose open research questions.
	With this contribution, we aim to provide the groundwork necessary for further improvements of power grid visualizations.

	\section{Related Work}
	\label{sec:related_work}
	The research field on power grid visualization is relatively sparse. We first present the history of the field, before looking at related visualization domains.
	
	\subsection{History of Power Grid Visualizations}
	The works of Guimaraes et al.~\cite{Guimaraes.2016} and also Nga et al.~\cite{Nga.2012} provide a quick overview over parts of the field:
	As mentioned in Mikkelsen et al.~\cite{MikkelsenC.2012}, few solutions have been proposed in the last two decades for visualizing control information for power grids. Pioneering work was done by Mahadev et al.~\cite{Mahadev.1993}. They were one of the first to realize that \emph{existing representational methods do not provide quick and efficient communication of the qualitative information contained}~\cite{Mahadev.1993} in the enormous amount of information contained in power grid data and no \emph{adequate representations have [...] yet been developed}~\cite{Mahadev.1993}.
	
	From an overall perspective, the first relevant work in the field of network visualization is generally considered to be by Becker et al.~\cite{Becker.1995}, where he is one of the first to focus on visualizing the data associated with the network (using a software called SeeNet) and not only the geographical network itself.
	One of the most important contributors to the field of power grid visualization in particular, especially in the (early) 2000s, but also, later on, was Thomas J. Overbye.
	He presents many novel and alternative approaches~\cite{Overbye.2000a, Overbye.2000b, Weber.2000, Overbye.2001, Overbye.vis.2001, Overbye.2003, Overbye.2005a,  Overbye.2005b, Overbye.2007, Overbye.2009, Overbye.2014}.
	Geospatial visualization methods for the transmission grid proposed in the literature are contour plots~\cite{Overbye.2000b, Overbye.2009, Overbye.2014}, Geographic Data Views~\cite{Overbye.2007}, %
	and a geospatial distortion approach called 'GreenGrid'~\cite{Wong.2009}.
	As mentioned in~\cite{Nga.2012}, Powerworld Simulator~\cite{Overbye.2000b} and General Electric's e-terra platform are tools widely used in the commercial sector, although the level and types of visualization they provide are somewhat behind the state-of-the-art.
	Research has been done on using the third dimension in combination with a two-dimensional diagram to visualize additional data. While Sun et al.~\cite{Sun.2004} did basic visualizations on line flow, the more advanced work of Milano et al.~\cite{Milano.2009} uses 3D plots for voltage magnitude and power flow while drawing the grid on the surface plot layer. Additional work on power flow was done by Boardman et al.~\cite{Boardman.2010}, but according to~\cite{MikkelsenC.2012, Nga.2012} up to 2012 and confirmed by~\cite{Cuffe.2015} in 2015, surprisingly few new ideas were proposed.
	A few approaches~\cite{rao.cn.2004, zhao.2014} were also considered in the Chinese-speaking community, like 3D power flow mappings in virtual reality.
	Indeed, using visualization and \emph{visual analytics to maintain situational awareness in the context of critical infrastructure is an active topic of research}~\cite{Jaeger.2016}.
	
	\subsection{Related Visualization Domains}
	\label{sec:methodology}
	A \emph{power grid } is an interconnected network of electrical transmission lines and associated equipment distributing electricity from electricity sources to consumers.
	Commonly, the vertices (e.g., plants, substations, transformers) are referred to as \emph{buses} (or a bus in singular), while the \emph{lines} refer to the connections between them.
    As such, it can be represented as a topological graph structure, with many associated (static and dynamic attributes like load, status, actual voltage, power consumption, flow, efficiency, or predicted demand, to just name a few.
    Simultaneously, the grid itself follows a geographic layout.
    This makes power grid visualization a hybrid between (dynamic) graph and geographic information representations.
	Accordingly, the central aim of power grid visualization is to search for ways to visualize the connectedness of the grid itself while retaining some geographic resemblance but simultaneously combining it with a data representation of one or several of the parameters mentioned above.
	This is done to find a meaningful representation of the system as a whole in an (interactive or dynamic) visualization. This visualization and the interaction methods should help users with operating, monitoring, and planning the network and should enable them to spot key issues and improvements that otherwise would be much harder to detect or even overlooked.
	
	\section{Survey}
	\label{sec:survey}
	
	In the following Section~\ref{sec:survey_overview} we give an overview of standard approaches and the most promising novel techniques proposed in the literature. 
	In Section~\ref{sec:survey_categorization} we propose classification methods and different criteria to assess the presented papers.
	A quick, qualitative overview is given in Table~\ref{tab:approaches_comparison}, while an in-depth comparison can be found in Section~\ref{sec:survey_comparison}.
	
	\subsection{Overview of current approaches}
	\label{sec:survey_overview}
	In the following, the approaches are ordered by their \emph{complexity} (detailed in Section~\ref{sec:survey_categorization}).
	For reference, we discuss two widely-used (generic) approaches: \emph{geographic} and \emph{topological} representations:
	
	\textbf{Reference: Geographic} --- %
	As one of the oldest forms of network display, the grid is shown on a geographic map based on the geospatial positions of its elements, either as accurate as possible (true mapping) or in a stylized and more abstract form. Note that there is a whole field concerned with the details of geographic visualizations.
	The geographic representation is preferred for highlighting geographical aspects of the network, and local features are represented. Examples include identifying interrupted lines, finding exact coordinates, or street-level planning.
	However, it does \emph{not necessarily represent the most desirable visualization because it fails to show the electronic view of an electric power system}~\cite[p.~412]{Wong.2009}.
	
	\textbf{Reference: Topological} --- %
	A classical topological network view represents the connectedness of the network. In general, topology and geography are partly correlated, as the physical infrastructure is related to some geographic context.
	Showing the interconnectedness supports identifying problems that might spread to neighboring grid elements, like a local power outage that could cascade through the network.
	While a strictly topological view, in general, does not scale well with increased network complexity and size, topological approaches can be used to reduce the network complexity and \emph{propagate incomplete information from neighbors}~\cite{Wiegmans.2016}. This is necessary for modeling because often, complete electrical data is only available for a small subset of the lines and stations, and much of the information has to be inferred (cf.~\cite{Wiegmans.2016}).
	Using topological structures instead of geographical representations is a major area for further research, and studies show that often the topological information is more relevant, but it might be helpful to retain (some) relative geographic orientation (cf.~\cite{Steiger.2013}).
	While not a topological view in the classical sense (showing simply the interconnectedness of the network), approaches that combine topology and geographic aspects and often incorporate the underlying physics of the network have the most potential for visualization power grids and are the main focus of this survey.
	
	\textbf{Traditional: Single Line Diagram} --- %
	\begin{figure}
		\centering
		\includegraphics[width=.95\columnwidth]{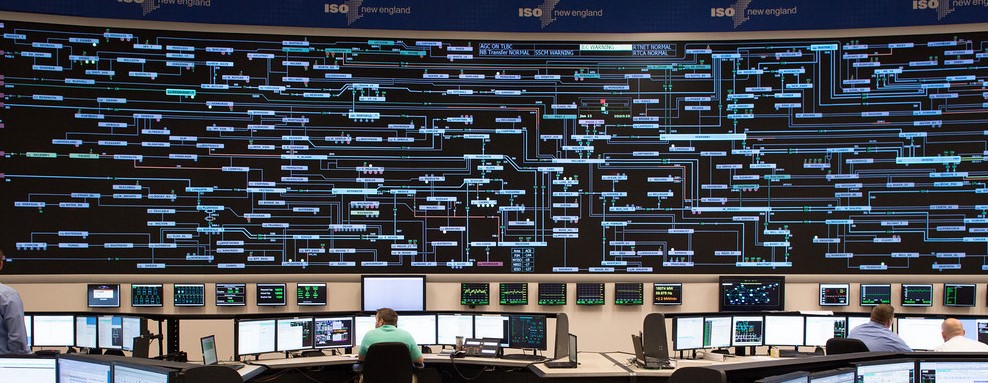}
		\caption{Single line diagram view~\cite{img.topologicalview.isonewengland}.}
		\label{fig:single_line_diagram}
	\end{figure}
	Single line diagrams are a basic variant of topological representations with some additional parameters displayed along the lines and stations, often using color, text, and values, like the one in Figure~\ref{fig:single_line_diagram}. Mostly these are limited to operating status and simple parameters. It is primarily used for a general overview of the network and only helps to detect simple network problems.%

	\textbf{Traditional: Time Series Bar Charts / 3D surface contour} ---%
	Bar charts %
	are a well-known technique to present data. While static displays show the current value and adapt as necessary, time series bar charts show the historical flow of a quantity, often updating the values in inverse reading direction. This can be used to detect changes and patterns in a single parameter over time.
	It is also possible to use a representation of times series data with 3D surface contour plots~\cite{Nga.2012}.
	
	\textbf{Scatter-Plot Matrix} --- %
	Scatter-plot matrices can be used to analyze medium-dimensional data (not more than ten to 15 dimensions).
	The diagonal axis can also show a histogram for each variable (instead of the dimension name). Often used as a step in data mining, they are very good at detecting bi-variate patterns or relationships between the dimensions and data clusters and outliers, while higher-dimensional structures are harder to detect. 
	
	\textbf{Parallel Coordinates / Andrews Curve} ---%
	To detect relationships among multivariate data, parallel coordinates can be used. Instead of using orthogonal axes, this plot uses a cartesian plot with all the dimensions spaced horizontally apart. This means the x-axis contains all the dimensions, while the y-axis all the data of each dimension. One data vector is then represented as a connected line over the whole width of the plot. Due to different ranges, the data is usually normalized.
	An alternative is the Andrews curve, a smoothened version of a parallel coordinate plot, with the data as parameters in a Fourier series.

	\textbf{Power / Line Flow} ---%
	Visualizing the current flows dynamically on a power grid view is described in~\cite{Overbye.2000a}. Arrows are superimposed over the lines and animated to visualize the current flow. According to the authors, \emph{a user can gain deep insight into the actual flows occurring on the system [at a glance]}. This supports familiarization and outlier detection.
	
	\textbf{Third dimension above grid plane} ---%
	Using the third dimension to represent additional data was first proposed by Sun et al.~\cite{Sun.2004}. More advanced work was then done by Milano et al.~\cite{Milano.2009} for the electrical properties of a grid.
	The general idea is to enhance the display of 3D physical data by using a representation similar to common 3D surface plots but substituting the surface layer for a grid view of the system. Using this approach, one can plot, for example, voltage magnitude and power flow and see the associated grid elements directly.
	
	\begin{figure}
		\centering
		\includegraphics[trim={0 0.1cm 0 1.8cm},clip,width=\columnwidth]{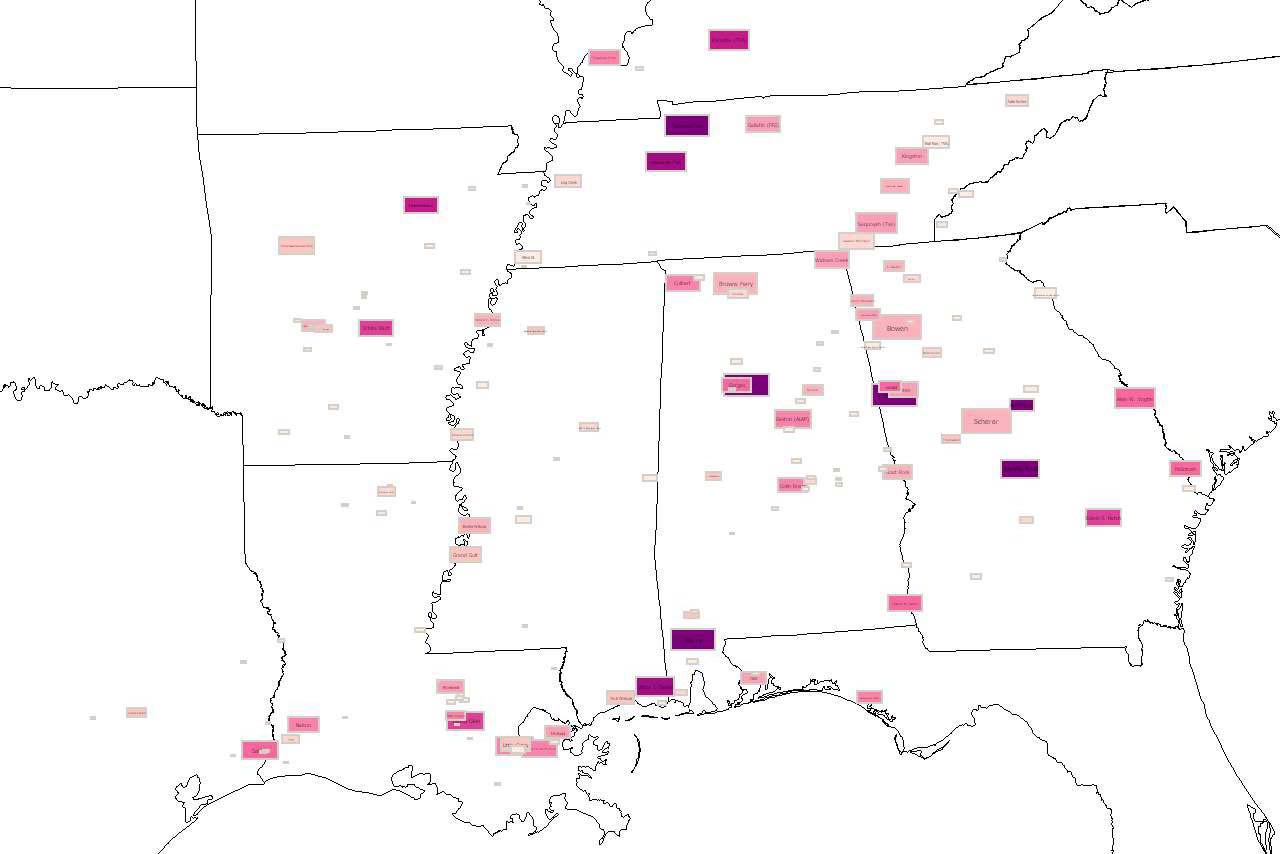}
		\caption{Geographic Data View where the size is proportional to generation and the fill color to reactive reserves~\cite{Overbye.2007}.}
		\label{fig:geographic_data_view}
	\end{figure}
	
	\textbf{Interactive 3D Visualization} ---%
	Overbye~\cite{Overbye.2000a} also proposed the use of 3D-CGI representations by levering the human's familiar perception with a three-dimensional world. So far, the only example includes representing bar charts as 3D objects superimposed over a 2D map. However, this approach is viable in Virtual Reality environments, for example, for maintenance and virtually displaying parts or work steps in a power grid station.

	\textbf{Geographic Data Views (GDV)} ---%
	This approach was proposed by Overbye~\cite{Overbye.2007}.
	It describes dynamically created visualizations, an example shown in Figure~\ref{fig:geographic_data_view},
	 which show a relevant subset of the network.
	The idea originates a predefined one-line format, which can be effective for monitoring due to consistency, and zooming can be used to get more detail. But for corrective control or analysis, it is problematic to design a priori views which encompass all the information needed in a particular scenario for effective decision making. This is also relevant in planning when aiming at a specific power flow pattern and deciding which lines to retain and which to connect. (cf.~\cite{Overbye.2007}). Such scenarios, for example, occur during the planning of Suedlink power line corridor.
	Geographic data view visualizations combine a power system with a geographic model and represent system parameters using graphical symbols. Thereby the location, size, color, rotation, and shape are dependent on the power system data (cf.~\cite{Overbye.2007}). The placement of information on a map is a primary research area in cartography.
	
	\textbf{Contouring} ---%
	\begin{figure}
		\centering
		\includegraphics[trim={0 1.2cm 0 0.4cm},clip,width=\columnwidth]{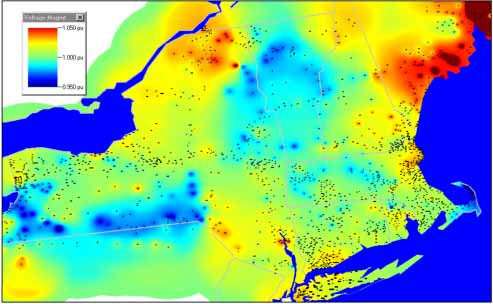}
		\caption{Contouring voltage on the US East Coast~\cite{Overbye.2000a}.}
		\label{fig:contouring}
	\end{figure}
	Contouring, presented in various forms \cite{Overbye.2000a, Overbye.2000b, Weber.2000, Overbye.2003}, like shown in Figure~\ref{fig:contouring}, describes displaying data on a given (often) geographical or (sometimes) topological representation using a colored contour graph plot, similar to a heat map. These maps can be displayed either as static or with dynamic elements.

	\textbf{GreenGrid} ---%
	\begin{figure}[b]
		\centering
		\includegraphics[trim={0 0.2cm 0 2.2cm}, clip, width=\columnwidth]{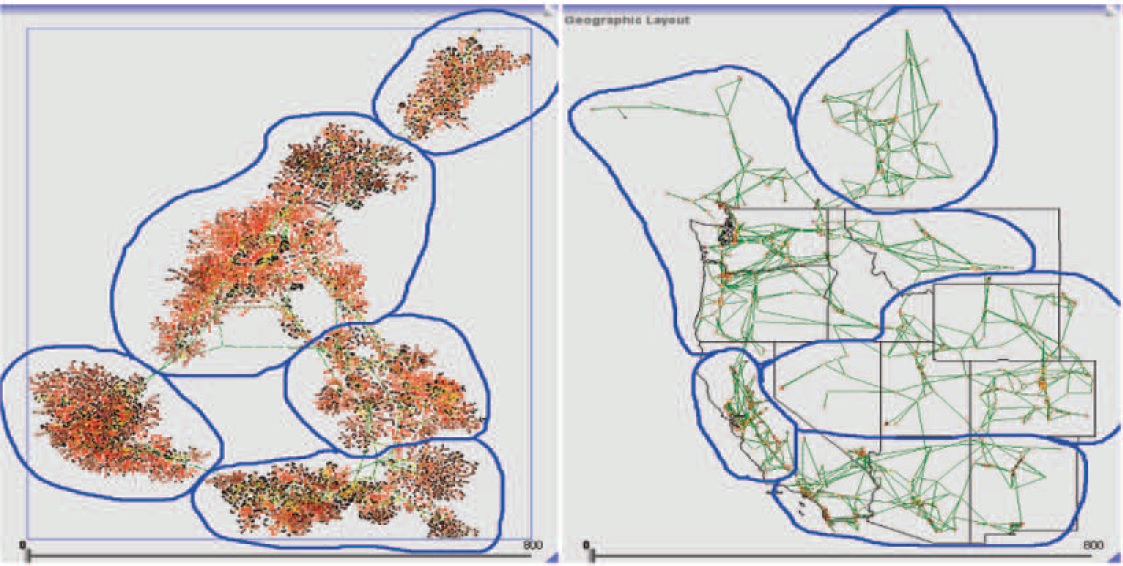}	
		\caption[GreenGrid visualization (left) according to voltage level compared to a geographical representation (right) with synchronous island circled of the WECC power grid~\cite{Wong.2009}]{GreenGrid visualization (left) according to voltage level compared to a geographical representation (right) with synchronous island circled~\cite{Wong.2009}.}
		\label{fig:greengrid}
	\end{figure}
	GreenGrid is a geospatial distortion approach using weighted force-directed visualizations to combine geographic views and transform them into topological representations with respect to some system parameters.
	It was proposed~\cite{Wong.2009} by Wong in 2009. The main advantage over other techniques is that one can visually represent the effect one or several parameters have on the grid. An example is given
	in their paper %
	where parts of the grid on the U.S. West coast are visualized according to their voltage levels. The comparison to the sole geographical view shows that the existence of clusters and their few interconnections become much clearer at a single glance. In the figure, synchronous islands are circled, where the network frequency is kept stable and synchronous. An accompanying usability study found that \emph{GreenGrid layout generally offers equal or improved performance over the geographic layout in terms of accuracy, time to completion, and user satisfaction for most problems in head-to-head comparison [while weaknesses are] its inability to represent and convey boundaries and areas}~\cite{Wong.2009}.

	\textbf{GreenCurve} ---%
	\begin{figure}
		\centering
		\includegraphics[trim={0 2.0cm 0 0.8cm}, clip,width=\columnwidth]{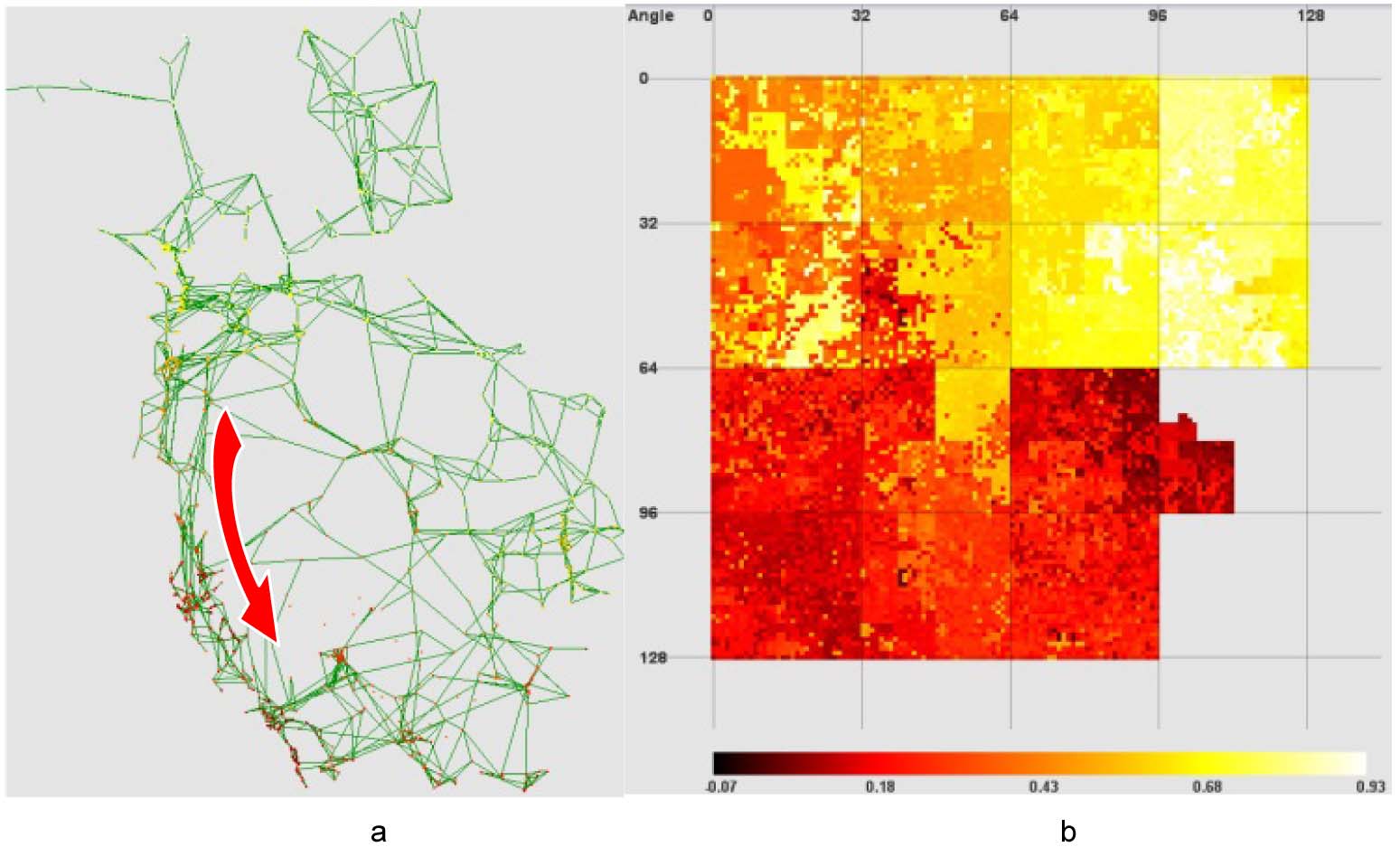}	
		\caption{GreenCurve visualization (right) compared to graph layout (left)~\cite{Wong.2012}.}
		\label{fig:greencurve}
	\end{figure}
	GreenCurve is a dense pixel display, the approach based on a space-filling two-dimensional Hilbert curve for displaying small multivariate graphs. It was proposed by the same authors~\cite{Wong.2012} as GreenGrid and can be seen as a complementary (but different) approach. The layout is denser, which is archived by solving the Fiedler vector of the Laplacian and thereby sorting neighboring vertices and then folding them into a fractal (Hilbert) curve (cf.~\cite{Wong.2012}). An accompanying usability study states that \emph{participants have found that they can solve critical network analysis problems more accurately and with higher satisfaction than using a force-directed graph layout [like GreenGrid]. The controlled experiment also showed that the GreenCurve and the force-directed graph layouts are highly complementary for specific kinds of network problems}~\cite{Wong.2012}.

	\textbf{Other approaches} ---%
	Other approaches found in literature \cite{zhao.2014} are virtual reality 3D power flow mapping, the use of splicing to display the whole grid in one view, using Gantt charts to visualize maintenance schedules in tabular-like visualization, k-lines for displaying load curves, the use of 3D visualizations for either synthetic alarms or low-frequency oscillations.
	
	\textbf{Commercial products} ---%
	According to~\cite{Nga.2012} and others, the two most common visualization products used in the industry are \emph{Powerworld Simulator}~\cite{Overbye.2000b} and General Electrics' \emph{e-terra platform}. Powerworld Simulator, based on research done by Overbye, uses, among others, geographical views, single-line views, time-series charts, contouring, and Andrews curves, while e-terra uses, among others, geographical views, single-line views, and time-series charts. In this regard, Powerworld Simulator contains more advanced visualization features. However, while it implements some of the more basic visualization techniques, it falls short on leveraging the state-of-the-art techniques proposed in academia.
	
	\subsection{Categorization and Assessment}
	\label{sec:survey_categorization}
	We first want to propose possible classifications and criteria to compare the different approaches.
	We will establish the following nomenclature to classify the techniques into three different categories in terms of \textbf{complexity}: simple (\symLow), intermediate(\symMedium), and advanced (\symHigh).
	
	Traditional techniques have been employed for some time (i.e., decades) in power grid visualization, for example, bar charts.
	We label techniques simple when the approach is a simple plug-in working solution of a well-known technique without many adaptions for power grids but has been proposed in the literature regarding power grid visualization.
	Intermediate techniques are often a combination of well-known graph or data representation approaches. Often, they show a graphical representation enhanced by symbols, animations, or heatmaps (contour plots). Main aspects are often \emph{how} to display the information, in which shape, color, form, size, having different views. They all have in common that they are stitched together from existing approaches and can, for a big part, be seen as a (large) set of possible view forms. They do \emph{not} contain \emph{new} or \emph{complex} algorithms to transform the visualization itself.
	Advanced methods encompass the more advanced ideas proposed, which would typically be considered state-of-the-art. They mainly deal with new visualization techniques that do not resemble usual power grid representations. They mostly build on existing work in graph drawing, mathematics, or other, too, but they can be set apart from the intermediate approaches in the way that a \emph{significant} effort has been put into adapting these approaches to the specific domain.
	
	As mentioned in~\cite{Nga.2012},~\cite{Cai.2009} proposed a useful classification of the \textbf{visualization techniques} in three categories: 
	low-dimensional data, multivariate high-dimensional data and Graph (geographic information) based visualizations.
	 Examples of low-dimensional approaches are simple 2D charts and 3D surface plots with contour. Multivariate high dimensional are, for example, parallel coordinates and scatter matrix plots.
	 We will follow his approach, but also consider time-series and interaction support.
	 Support for \textbf{time-series} analysis stands in contrast data to a steady-state visualization.
	 In terms of \textbf{interaction}, one can differentiate if a method is primarily static or provides manipulation methods, for example, by setting different parameters.
	We distinguish between the approaches according to four key properties:
	The main \textbf{algorithm} employed, the use of \textbf{highlighting} or pictograms to display data, \textbf{real time} support (computation power), and if the approach makes use of the actual \emph{physics} of the electric grid to present an appropriate visualization.
	The last aspect to differentiate the papers and approaches is the presence or absence of a usability or case \textbf{study}.
	
	\subsection{Comparison}
	\label{sec:survey_comparison} 
	
	\begin{table*}
		\caption{Comparative overview of the visualization techniques for power grids. For a detailed discussion, see Section~\ref{sec:survey_comparison}.}
		\label{tab:approaches_comparison}
		
		\centering
		\renewcommand{\arraystretch}{1.3}
		\setlength{\tabcolsep}{9.0pt}
					
    	\fontsize{7}{8} \selectfont
		
		\begin{tabular}{l|l|D|AAAAA|BBBB|C}
		    
		    \toprule
		    
			\multicolumn{2}{c|}{Paper} &
			\multicolumn{1}{D|}{Method} &
			\multicolumn{5}{A|}{Visualization}& %
			\multicolumn{4}{B|}{Properties}  &
			\multicolumn{1}{C}{Study} \\
			
             \cmidrule{1-2}  \cmidrule{3-3} \cmidrule{4-8} \cmidrule{9-12} \cmidrule{13-13} 
			
			\rotatebox[origin=l]{90}{Approach} &
			\rotatebox[origin=l]{90}{Reference} &
			\rotatebox[origin=l]{90}{Complexity} &
			\rotatebox[origin=l]{90}{Low-dim.}&
			\rotatebox[origin=l]{90}{High-dim.} &
			\rotatebox[origin=l]{90}{Graph} &
			\rotatebox[origin=l]{90}{Time-Series} & %
			\rotatebox[origin=l]{90}{Interaction} &
			\rotatebox[origin=l]{90}{Algorithm} &
			\rotatebox[origin=l]{90}{Highlighting} &
			\rotatebox[origin=l]{90}{Real-time} &
			\rotatebox[origin=l]{90}{Physics} &
			\rotatebox[origin=l]{90}{Evaluation} \\
			
			\midrule

			Time-Series Bar Chart &
			\cite{Overbye.2000a} &
			\symLow &
			\symYes &
			\symNo &
			\symNo &
			\symNo &
			\symNo &
			{geospatial} &
			{symbols} &
			\symYes &
			\symNo &
			\symNo \\
			
			Power flow &
			\cite{Overbye.2000a} &
			\symLow &
			\symNo &
			\symNo &
			\symYes &
			\symYes &
			\symNo &
			{geospatial} &
			{symbols} &
			\symNo &
			\symNo &
			\symNo \\
			
			3D Surface &
			\cite{Milano.2009} &
			\symMedium &
			\symYes &
			\symNo &
			\symYes &
			\symYes &
			\symNo &
			{surface plot} &
			- &
			\symPartial &
			\symNo &
			\symNo \\
			
			Interactive 3D &
			\cite{Overbye.2000a} &
			\symMedium &
			\symYes &
			\symNo &
			\symYes &
			\symYes &
			\symYes &
			{geospatial} &
			{symbols} &
			\symPartial &
			\symNo &
			\symNo \\
			
			VA-based &
			\cite{Mittelstaedt.2013b} &
			\symMedium &
			\symYes &
			\symNo &
			\symYes &
			\symYes &
			\symNo &
			geospatial  &
			{symbols} &
			\symYes &
			\symNo &
			\symNo \\

			Contouring & 
			\cite{Overbye.2000a}$^\ddagger$ &
			\symMedium &
			\symYes &
			\symNo &
			\symYes &
			\symYes &
			\symNo &
			{graph / geospatial} &
			{color} &
			\symNo &
			\symYes &
			\symNo \\
			
			Geographic Data View & 
			\cite{Overbye.2007} &
			\symHigh &
			\symYes &
			\symNo &
			\symYes &
			\symYes &
			\symYes &
			{geospatial} &
			{symbols} &
			\symNo &
			\symNo &
			\symYes \\
			
			GreenGrid &
			\cite{Wong.2009} &
			\symHigh &
			\symYes &
			\symYes&
			\symYes &
			\symYes &
			\symYes &
			{force-directed} &
			{color} &
			\symYes &
			\symYes &
			\symYes \\

			GreenCurve & 
			\cite{Wong.2012} &
			\symHigh &
			\symYes &
			\symYes&
			\symYes &
			\symYes &
			\symYes &
			{fractal curve} &
			{color} &
			\symYes &
			\symYes &
			\symYes \\
			
			\midrule
			
			e-terra{$^\dagger$} & 
			- &
			\symMedium &
			\symYes &
			\symNo  &
			\symYes &
			\symYes &
			\symYes &
			geospatial &
			symbols &
			\symYes &
			\symPartial &
			\symPartial \\
			
			PowerWorld$^\dagger$ &
			\cite{Overbye.2000b} &
			\symMedium &
			\symYes &
			\symNo &
			\symYes &
			\symYes &
			\symYes &
			geospatial &
			symbols &
			\symYes &
			\symPartial &
			\symPartial \\
			
			\bottomrule
			
			\multicolumn{12}{>{\linespread{0.95}\scriptsize}p{13cm}}{%
			\textbf{$\dagger$}: commercial approaches $\qquad$  $\ddagger$: and later works: \cite{Overbye.2000b, Weber.2000, Overbye.2003} $\qquad$
			\textit{Legend}: $\;$
			    {-}~None $\,$
				\scalebox{0.8}{\symNo}~Missing $\,$
				\scalebox{0.8}{\symPartial}~Partly $\,$
				\scalebox{0.8}{\symYes}~Present
		} \\
		\end{tabular}
	\end{table*}
	
	A qualitative comparison of the important features from the most promising approaches identified in Section~\ref{sec:survey_categorization} is given in Table~\ref{tab:approaches_comparison}. This table can also provide a quick overview of the main findings of this survey. Sometimes the labels given in the table are over-simplified and do not give the whole aspect credit. The complete evaluation is provided in the following sections.
	The criteria for comparison have been explained in Section~\ref{sec:survey_categorization}. In the following, we compare the features of the individual approaches.
	
	\textbf{Complexity} ---%
	For reference, we presented geographical and topological representations. Traditionally, single-line diagrams, time-series bar charts in either 2D or 3D have been used. Simple approaches (\symLow) are the use of well-known techniques like scatter-plot matrices, parallel coordinates, or the introduction of animated flows in a power flow representation. Intermediate complexity (\symMedium) applies to techniques with moderate adaptions, like 3D surfaces, Interactive 3D Visualization, VA-based, and Contouring.  Advanced techniques (\symHigh) use complex algorithms for information placement or rework existing technologies to a larger degree, like Geographic Data Views, GreenGrid, and GreenCurve.
	
	\textbf{Visualization technique} ---%
	All presented approaches - with the exception of times series bar charts - are based on geographical (graph) information. While time series bar charts present only low-dimensional data, 3D surfaces, Interactive 3D Visualization, VA-based, and Geographic Data Views, as well as Contouring, present low dimensional on top of geographical information. This is different for GreenGrid and GreenCurve, whereby adapting the weight functions, multi-dimensional data, and relationships can be incorporated. This makes them, apart from scatter-plot matrices and parallel coordinates, the only approaches that can display multivariate high-dimensional data.
	
	\textbf{Time-Series Data} ---%
	Geographical and simple topological views are static by definition, as long as the elements themselves (their location or their connectedness) do not change. Single-Line-Diagrams can show time-series-data in a basic way (e.g., normal operations, critical, off-range), while time-series bar char do a better job at displaying univariate information with a time component. Scatter-Plot matrices and parallel coordinates in their normal variants are unable to display time-series data of arbitrary granularity and are better not used at all for such a task. Power flow, 3D surfaces, Interactive 3D Visualization, VA-based, Geographic Data Views, Contouring, GreenGrid, and GreenCurve, are all able to display times series data in (near) real-time, although performance can be negatively impacted.
	
	\textbf{Interaction} ---%
	Again, in their basic form, geographical and topological representations are both static. Pan and zoom are disregarded when the view is just zoomed and not updated with new or more detailed information on the fly.
	The same is in principle valid for single line diagrams, although variants exist that do provide some interactivity. Geographic Data Views can be seen as such an extension. Scatter-plot matrices and parallel coordinates are, however, strictly static. Power flow, 3D Surfaces, and Contouring are considered static, as they just display the dynamic data but cannot be interacted with. Visual Analytics, Interactive 3D Visualization as well as Geographic Data Views, however, can be seen as strictly interactive, as their main goal is to provide custom and interactively adapting views. GreenGrid and GreenCurve also belong to this category, as the parameters for the visualization are adjusted interactively and dynamically; also, some pre-computed (static) parts exist.
	However, boundaries between static and interactive are not strict, and overlapping can exists. Also, some of the techniques could be adapted to provide interactivity.
	
	\textbf{Algorithms} ---%
	Geographical views are commonly plotted on a 2D surface using a reference ellipsoid (like WGS84) to map geo-coordinates to a 2D map. Topological representations draw heavily from research in the field of graph drawing, where numerous approaches exist.
	The other approaches mentioned mostly rely on geographical views that are dynamically enhanced by adding additional information, like geospatial animations on a map for Power flow or symbols for Geographic Data Views. Contouring uses contour-based surface plots on top of geospatial data. GreenGrid uses physical properties to adapt the weights of a force-directed graph in a non-trivial way to generate a mixture of a topological and geographical representation, distorted according to the physics of the weight function and parameter used. GreenGrid has the most involved approach, being based on a space-filling fractal (Hilbert) curve. To reduce the size and compact the representation, the adjacent nodes are ordered accordingly, using spectral graph partitioning. For this, the second smallest eigenvector of the graph Laplacian is used. The resulting nodes are then folded into a fractal curve.
	
	\textbf{Highlighting} ---%
	Geographical, Topological representations, and scatter plot matrices do not use any highlighting by default. Single-Line-Diagrams can use symbols and time series bar charts, as well as parallel coordinates, can use color to distinguish between data points. Most approaches that deal with interactive views use symbols of various size, color, location, rotation, and shape, that is dependent on the power system data.
	Contouring uses color when displaying contour graphs (heatmaps). This is also true for GreenGrid and GreenCurve, where both use color to encode physical features.
	
	\textbf{Real-Time} ---%
	For all the algorithms presented, memory demands are no issue. While Interactive 3D Visualization arguably needs the highest amount to create a 3D world, due to the expertise in VR research and computer game development, efficient methods exist. Due to the comparably small size of single power grid networks (less than one million nodes, typically), geographical and graph drawing can be done using algorithms developed for analytical network mining in social or biological research. Also, problems like cluttering or occlusion may occur; techniques like edge bundling, clustering, and others exist to cope with these problems. Computation power is of no issue for all presented methods if efficient algorithms are used. The studies for GreenGrid~\cite{Wong.2009} cite an average computation time of around 1/10th of a second in 2009 and for GreenCurve~\cite{Wong.2012} 2012 of not more than two seconds, which can be considered near-real-time.
	
	\textbf{Physics} ---%
	Of the approaches presented, all show some aspect of the physics involved, but most are trivial, for example, using either the location or connectedness, showing plain values, or the current direction.
	Contouring is somewhat more involved as it is able to overlay physical attributes and trends over a network representation as heat-maps, which can provide a faster understanding of the physics involved.
	GreenGrid uses different physical properties to adapt the weights of a force-directed graph in a non-trivial way to help distinguish certain aspects. GreenCurve uses mathematical methods in physics and from classical mechanics and math to display physical properties of the network as a fractal curve, thereby creating an entirely new representation.

	\textbf{Study / Evaluation} ---%
	Geographical, Topological representations and Times Series Bar Charts have been in use for decades in the power grid industry. These representations are ubiquitous and provide the basis for other, more detailed views. Studies discussing the different approaches only exist for Geographic Data Views, GreenGrid, and GreenCurve. The authors from GreenGrid did domain expert studies and found that productivity~\cite{Wong.2009} and accuracy increased by up to a factor of two. However, due to the different viewing paradigm, the accuracy for border detection was reduced. A study of GreenCurve~\cite{Wong.2012} showed that the complex viewing paradigm alone could sometimes lead to reduced productivity and accuracy. This, however, is mitigated when combining GreenGrid and GreenCurve representations, where accuracy and productivity increased up to a factor of two compared to GreenGrid alone. Studied tasks include identifying transmission lines, phase angles, synchronous and asynchronous islands, ranking of node population, identifying states as energy sources, conduit or sinks,  and ranking of states by node population (cf.~\cite{Wong.2009}).
	
	\textbf{Usage / Applications} ---%
	All techniques are relevant and can be used during operations and monitoring of the network. Approaches like power flow and Contouring are relevant for market-based decisions like buying/selling power. In case of finding anomalies (like signs for an impending black-out due to dis-synchronized phases), approaches like Contouring and especially GreenGrid and GreenCurve are most promising. Their relevance in such a scenario has been shown in the accompanying studies. The same can be said for planning, especially in combination with simulation, where edge cases and different network layouts are tested.

    \subsection{Findings}
	In many cases, especially for new techniques, the approaches were untested, and the paper comments on usability studies as an area for further research. Secondly, most approaches deal only with the visualization of steady states. Sometimes~\cite{Nga.2012} the visualization of time-varying dynamic simulations is proposed for further research.
	Most papers begin with a quick and more or less comprehensive overview of related work.
	The specialized papers mostly acknowledge the shortcomings of current approaches and discuss the rationales underlying the development and design of the new approach. The next section describes the implementation, and sometimes the performance is assessed. Then the approach is compared in strengths and weaknesses to current (where the papers have different views of what constitutes current) geolocation-based visualization techniques for power grids. A few papers, however, also go on and present a usability~\cite{Weber.2000, Jaeger.2016} or case~\cite{Wong.2009} study. This is done by using either domain experts and measure their performance in fulfilling given tasks or assessing the visualization quality for issue detection (like a cascading black-out) using historical data in the latter case to evaluate the practical significance of the new visualization.
	
	\section{Discussion and Outlook}
	\label{sec:discussion}
	Visualization aims to provide an abstract representation to enhance and speed up understanding of complex data. This is especially true for power grid visualization, where the underlying physics introduces huge, high-dimensional data-sets, for which an efficient and useful representation often is not known a priori, and more often than not depends mainly on the specific problem that needs to be solved.
	The research question was successfully answered: We have shown which techniques are currently employed in the power grid industry, which is relatively simple and to a large degree dating back decades. This is surprising, given the technological advances in the last century, especially during the previous two decades. Reasons for this are discussed in Section~\ref{sec:outlook}. Furthermore, we have presented three more modern approaches, two of which fundamentally change how we look at the grid representation itself. For these approaches, we have cited accompanying studies that show how they significantly improve and simplify network analysis, with some example tasks given. The performance and accuracy can sometimes be doubled when solving the usual task during planning, operating, or managing a power grid (cf.~\cite{Wong.2012}).
	With the advent of smart grid structures, it becomes increasingly important to understand the inner workings of the grid. Moving from a centrally distributed delivery network to an inter-meshed network with millions of smart feeders and consumers introduces new challenges due to the added complexity and much more dynamic nature. These challenges come in the form of upgrading legacy designs, risk prevention in terms of critical infrastructure (see~\cite{Overbye.2014} for a discussion of this topic), (near) real-time monitoring and adaptation, all of which fuel the need for analytical insights for planning and operations. This can be provided and made possible by appropriate simulation and visualization methods.
	
	\label{sec:outlook}
	The importance of visualizing power grids will further increase in the years to come.
	The more it is surprising that no other approaches have been published and visual analytics techniques are rarely employed. This can be attributed either to a lack of research in this area in the last years, which just begins to increase again, or because some works are unpublished and directly integrated into or developed for commercial products, like the power grid simulation framework from Adpatricity~\cite{company.adaptricity}.
	A significant obstacle in this regard is \textbf{data unavailability}, either due to security or economic aspects. It concerns grid structure, load data, energy source locations, and specifications. Grid operators are rather protective of their data, and only some research projects were able to gain limited access to the data. Also, national transparency efforts resulted in some (basic) data being published (cf.~\cite{Medjroubi.2017}), but it remains a significant problem that slows research.
	From a technical perspective, visualizing the \textbf{dynamic nature} of the network in the form of several possible contingency states instead of a single snapshot is one of the core issues, along with \emph{human factor assessments}, and \emph{industry acceptance}~\cite{Overbye.2014}.

	\section{Conclusion}
	\label{sec:conclusion}
	This survey paper provides an overview of the latest methods developed for the visualization of power grids. The field is underdeveloped, and few significant publications exist. Many approaches directly use geographic or topological representations without adaptions. Some novel ideas have been proposed, like Geographic Data Views~\cite{Overbye.2007}, the use of 3D surface plots~\cite{Milano.2009}, weighted force-directed approaches~\cite{Wong.2009}, or space-filling fractal curves ideas~\cite{Wong.2012}, 
	The approaches under consideration have been compared according to criteria like complexity, visualization technique, algorithms, performance, use of highlighting, display of the underlying physics, the capability of displaying time-series data, interaction support, and evaluation. The techniques Contouring~\cite{Overbye.2000a, Overbye.2000b, Weber.2000, Overbye.2003} as well as  GreenGrid~\cite{Wong.2009} and GreenCurve by~\cite{Wong.2012} have been identified as the most advanced methods currently available; only the first is yet employed in commercial applications.
	With the emergence of smart grids, visualization and simulation plays an increasingly important role, helping to understand the capabilities and limits of a given network, identify and resolve specific issues and optimize the network structure to adapt to these new usage scenarios.
	We hope that this survey provides a basis for practitioners to leverage latest progress in different areas as graph drawing, dynamic network representation or cartography to support and improve visualization techniques for power grids in the mid- and long term.

	\bibliographystyle{eg-alpha-doi}
	
	\bibliography{bibliography}

\end{document}